\documentclass{ws-p9-75x6-50}

\newcommand{\lesssim}{\put(2,3){{\small $<$}}\put(2,-3){{\small $\sim$}}\quad}

\begin{document}

\title{Quantum Phase Transitions in Spin Systems}

\author{J.~Richter and  S.E.~Kr\"uger}

\address{Institute for Theoretical Physics, University of Magdeburg,
P.O.Box 4120, D-39016 Magdeburg, Germany\\E-mail:
johannes.richter@physik.uni-magdeburg.de}

\author{D.J.J.~Farnell and R.F.~Bishop}

\address{Department of Physics, University of Manchester Institute of
         Science and Technology (UMIST),
         P.O.Box 88, Manchester M60 1QD, United Kingdom}


\maketitle

\abstracts{
We discuss the influence of strong quantum fluctuations on zero-temperature
phase transitions in a two-dimensional spin-half Heisenberg system. 
Using a high-order coupled cluster treatment, we
study competition of magnetic bonds with and without frustration. We find that
the coupled cluster treatment is able to describe the zero-temperature
transitions in a qualitatively correct way, even if frustration is present
and other methods such as quantum Monte Carlo fail.}

\section{Introduction}
Phase transitions have been a subject of great interest to physicists over
many decades. Besides classical or thermal phase transitions, 
the so-called quantum phase transitions (or zero-temperature
transitions) have started to attract a lot of attention (e.g., see
Ref.\cite{qpt}). 
%To understand common, as well as different, aspects of thermal and 
%zero-temperature transitions it is useful to study simple models. 
The study of a variety of simple models allows us to understand which aspects
of thermal and zero-temperature phase transitions are common to classes of 
models and which are more special.
For continuous order-disorder transitions 
we basically need the interplay between the interparticle interactions and
fluctuations. For thermal transitions the Ising model may serve as the simplest
model. The equilibrium state corresponds to a minimum of the
free energy, and we have 
competition between energy and entropy controlled by the temperature.
For zero-temperature transitions no thermal fluctuations are present, and 
the fluctuations arise due to Heisenberg's uncertainty principle. A corresponding
basic model which has strong quantum fluctuations is the spin-half 
Heisenberg antiferromagnet (HAFM), particularly in low dimensions.

The subject of quantum spin-half Heisenberg
 antiferromagnetism in low-dimensional systems 
 has attracted a great deal of interest in
 connection with
 the magnetic properties of the high-temperature
 superconductors.
  Although we know from the Mermin-Wagner theorem\cite{mermin66} that 
 thermal fluctuations are strong enough to
 destroy magnetic long-range order (LRO) at any finite temperature in 1D and 2D, 
 the role of quantum fluctuations is less understood.
 It is now
 clear that the ground-state of the
 HAFM in 1D is not long-range ordered, whereas 
the HAFM on the square lattice is
long-range ordered (e.g., see Ref.~\cite{manou91}).
However, in 2D there are many other lattices with different coordination
numbers and topologies, and there is no general statement concerning
zero-temperature N\'eel-like LRO.

Anderson and Fazekas have
 suggested\cite{anderson73} that additional 
competition between magnetic bonds may increase quantum fluctuations 
and can suppress the N\'eel-like
 LRO in 2D. Indeed, the strength of this 
competition may serve as the
control parameter of a zero-temperature  order-disorder transition.
The competition between magnetic bonds in quantum spin systems 
can be caused in various 
ways.  As for classical spin systems, frustration can affect the magnetic
ordering in quantum spin systems. 
In the classical HAFM the frustration often
leads to canted (e.g., spiral) spin states which may or may not have counterparts
in the quantum HAFM. Furthermore, due to frustration 
Marshall's sign rule need not be fulfilled.\cite{MP_rule}
The violation of the sign rule in
frustrated systems makes their theoretical investigation particularly
difficult. For example, the quantum Monte Carlo (QMC) method 
suffers from the minus
sign problem in frustrated spin systems.

 A generic model of a frustrated HAFM is the spin-half
$J_1$--$J_2$ model on the square lattice,
 where the frustrating $J_2$ bonds plus quantum fluctuations yield 
a second-order transition from a N\'eel-ordered state to a disordered
quantum spin liquid
(see, e.g., Refs.~\cite{ri93,oitmaa96,Bishop6,sorella00}). 
On the other hand, there are examples where frustration leads to a
first-order transition in quantum spin systems in contrast to a second-order
transition in the corresponding classical model (see, e.g., 
Refs.~\cite{xian95,niggemann97,richter98,koga00}).

Besides frustration
there is a second type of competition between bonds which favours
a N\'eel-like distribution of spin
correlations over the lattice and other bonds which favour  
the formation of local spin singlets.
By contrast to frustration, which yields
competition in quantum as well as in classical systems, this type of
competition is present only in quantum systems. The formation of local
singlets is accompanied by the  `melting' of the magnetic 
LRO.
 This mechanism for breaking magnetic LRO may be relevant
 for the quantum disordered state in 
bilayer systems\cite{sandvik_scalapino,gros_wenzel_richter} as 
well as in CaV$_4$O$_9$ (see, e.g., Refs.~\cite{troyer96,troyer97}).
Of course both mechanisms can be mixed as, for instance, in SrCu$_2$(BO$_3$)$_2$
(see, e.g., Refs.~\cite{kageyama99,koga00a,koga00,richter98}).

In this paper we discuss the quantum order-disorder
transition driven by local singlet formation as well as 
the influence of quantum fluctuations on zero-temperature transitions 
driven by frustration. 
To that end we study a spin-half  model on the square lattice
in which both mechanisms,
frustration and singlet formation, are observed in
different regions.
High-order implementations of the coupled cluster method (CCM) are used to obtain
a consistent description of both types of competition for this model.
The CCM (see, e.g., Ref.~\cite{bishop91}) is 
one of the most powerful and most universal techniques in 
quantum many-body theory, and has previously been  applied
to quantum spin 
systems\cite{bishop91a,Bishop2,zeng98,Bishop6,bishop99,kruger00}
with great success.
In particular, we shall study to what extent the CCM is able to describe
zero-temperature transitions in spin systems. 

%%%%%%%%%%%%%%%%%%%%%%%%%%%%%%
\section{The Model}
%%%%%%%%%%%%%%%%%%%%%%%%%%%%%%
We consider a spin-half Heisenberg model on a square
lattice with two kinds of nearest-neighbour bonds $J$ and $J'$,
as shown in Fig.~\ref{fig1},
\begin{equation}\label{ham}
        H  =  J\sum_{\langle ij\rangle_1}{\bf S}_i\cdot{\bf S}_j
+J'\sum_{\langle ij\rangle_2}{\bf S}_i\cdot {\bf S}_j  
.\end{equation}
The expressions $\langle ij\rangle_1$, and $\langle ij\rangle_2$  
indicate nearest-neighbour 
bonds arranged in a regular zigzag pattern, as shown in Fig.~\ref{fig1} 
by the dotted and solid lines, respectively.
Each square-lattice plaquette
consists of three $J$ bonds and one $J'$ bond. If $J'$ and $J$ have 
different signs then the plaquettes are frustrated, whereas
competition without frustration is realized for antiferromagnetic bonds
$J'>J>0$. 
%For special values of $J$, $J'$ the model presents 
%(i) the Heisenberg antiferromagnet (ferromagnet) 
%on the square lattice for $J=J'=1$ ($J=J'=-1$), 
%(ii) the Heisenberg antiferromagnet (ferromagnet) 
%on the honeycomb lattice for $J=1$, $J'=0$ ($J=-1$, $J'=0$) 
%(iii) the spin one Heisenberg antiferromagnet on the triangular lattice 
%for $J=1$, $J'=-\infty$.
% KW

%%%%%%%%%%%%%%%%%%%%%%%%%%%%%%
\section{The Classical Ground State} \label{cgs}
\begin{figure}[t]
\epsfxsize=13.8pc % 12.7 spart eine Zeile ein will enlarge or reduce the postscript figures based on the xsize
\epsfbox{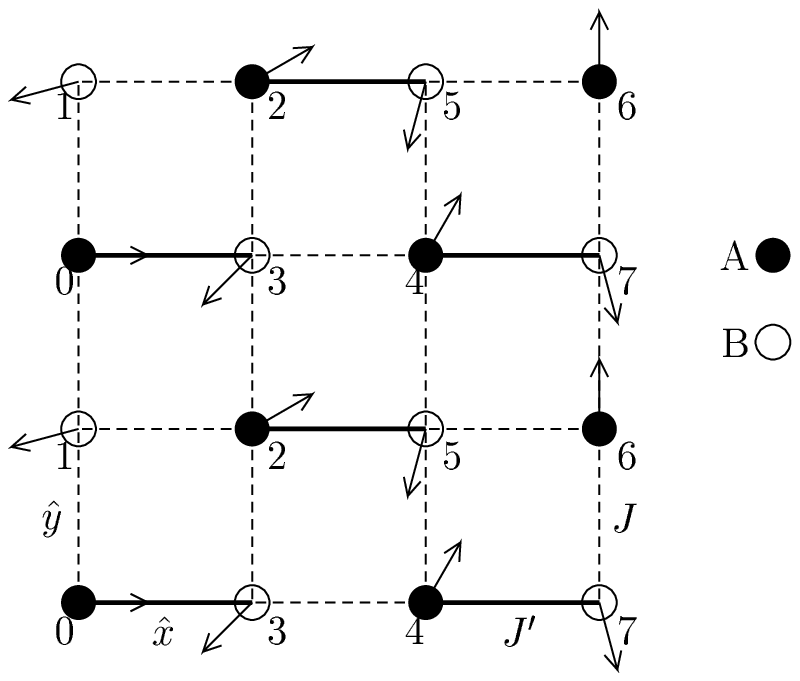} \epsfxsize=13.8pc \epsfbox{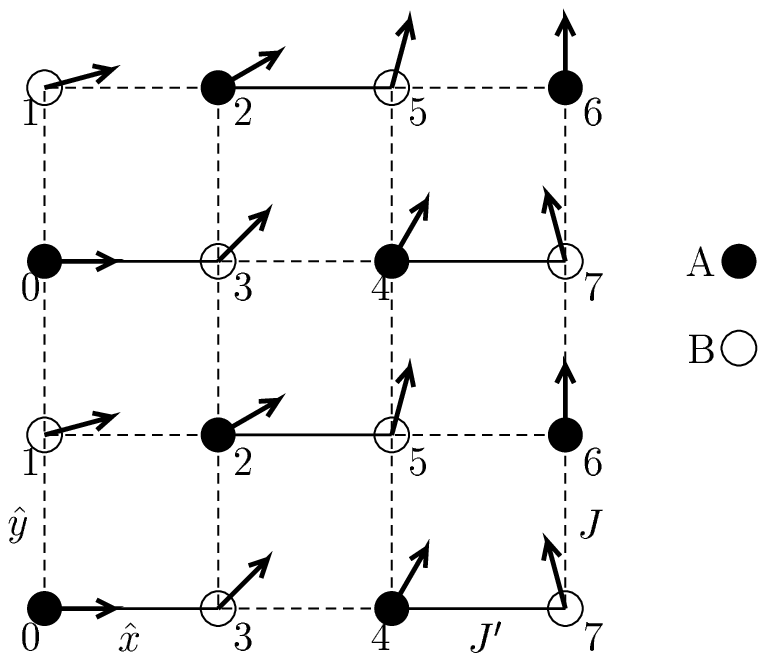} % postscript image file name
\caption{Illustration of the model ($J$-bonds correspond to dotted lines  and 
$J'$-bonds to solid lines) and of the classical spiral state for
antiferromagnetic $J=+1$ and ferromagnetic $J'<-1/3$ (left graph) and vice
versa ($J=-1$, $J'> 1/3$) (right graph).
As discussed in the text, both spiral states can be transformed into 
each other by reversing all of the spins on the B sublattice.
  \label{fig1}}
\end{figure}

%%%%%%%%%%%%%%%%%%%%%%%%%%%%%%
To discuss the influence of quantum fluctuations on the ground-state (GS)
properties of the model  we need to know the classical GS 
of  Eq.~(\ref{ham}).
%%(i.e. the ${\bf S}_i$ are classical vectors).
We set $J=1$.  Without frustration ($J'>0$),
the classical GS is the N\'eel state, independent of the strength
of $J'$. Frustration appears for 
(ferromagnetic) $J'<0$. One finds that the 
N\'eel state remains the GS for small amounts of frustration, $0 \ge J' \ge
-1/3$. At the critical point $J'=-1/3$, a second-order transition takes place
from the N\'eel state to a spiral state characterized by a pitch angle
$\Phi_{\rm cl}=\arccos(\sqrt{1-1/J'}/2)$ (see left graph of Fig.~\ref{fig1}).
We note that $\Phi_{\rm cl}=0$ (for $J'>-1/3$) corresponds to the N\'eel state.
The spin directions belonging to the $A$ and $B$ sublattices respectively, 
are given by 
$
{\bf S}_A({\bf R}) ={\bf e}_x\cos{\bf Q\cdot R}+{\bf e}_y\sin{\bf Q\cdot R}$
and 
  $   
     {\bf S}_B({\bf R}+\hat{x}) ={\bf e}_x\cos({\bf Q\cdot R}+\pi+3\Phi)
     +{\bf e}_y\sin({\bf Q\cdot R}+\pi+3 \Phi)$,
where 
${\bf R}$ runs over the sites of
the sublattice $A$, and  
${\bf Q}=(2\Phi,0)$.
We note that this spiral state is 
incommensurate in the $x$-direction.
% Next we consider ferromagnetic $J=-1$. 
By contrast to the quantum case, the classical model with $J=-1$
can be transformed into
the model with $J=1$ considered above by the simultaneous 
substitution $J \to -J$, $J' \to -J'$, ${\bf S}_{i \in B} \to -
 {\bf S}_{i \in B}$. 
Hence the physics for $J=-1$ is classically the same as
for $J=+1$ (c.f., Fig.\ref{fig1}).

\section{Competition Without Frustration} \label{nofr}
In this section we restrict our attention to the region where $J=+1$ and $J'>0$.
The terms within the Hamiltonian now compete because the first term 
favours N\'eel LRO  on the ``honeycomb'' lattice,
whereas the second term favours an uncorrelated product state of local pair 
singlets (see below). Again, we emphasize 
that there is no competition in the classical model. 

\begin{figure}[t]
\epsfxsize=12.0pc % will enlarge or reduce the postscript figures based on the xsize
\epsfbox{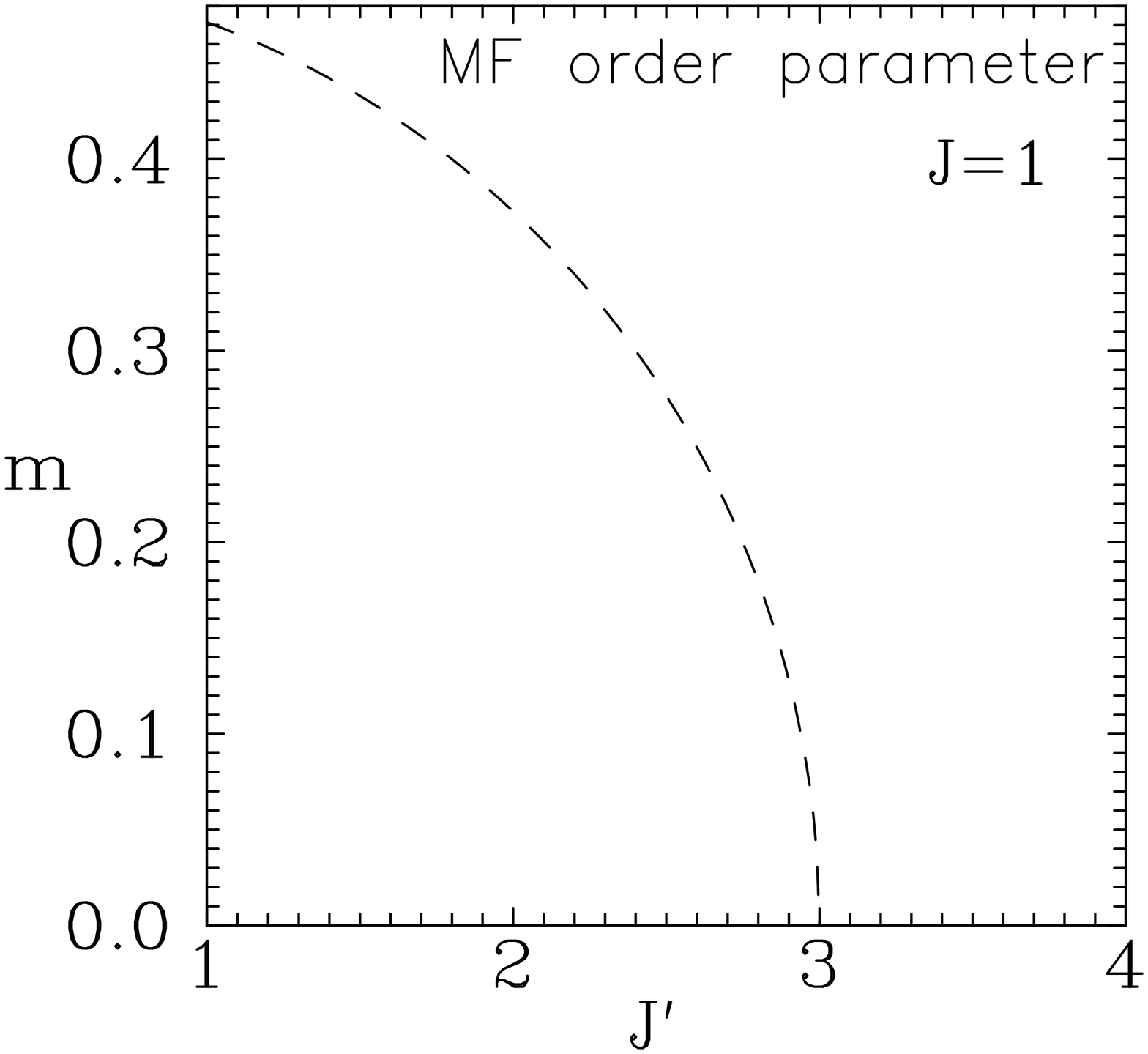}  % postscript image file name
\epsfxsize=12.0pc % will enlarge or reduce the postscript figures based on the xsize
\hspace{1cm}\epsfbox{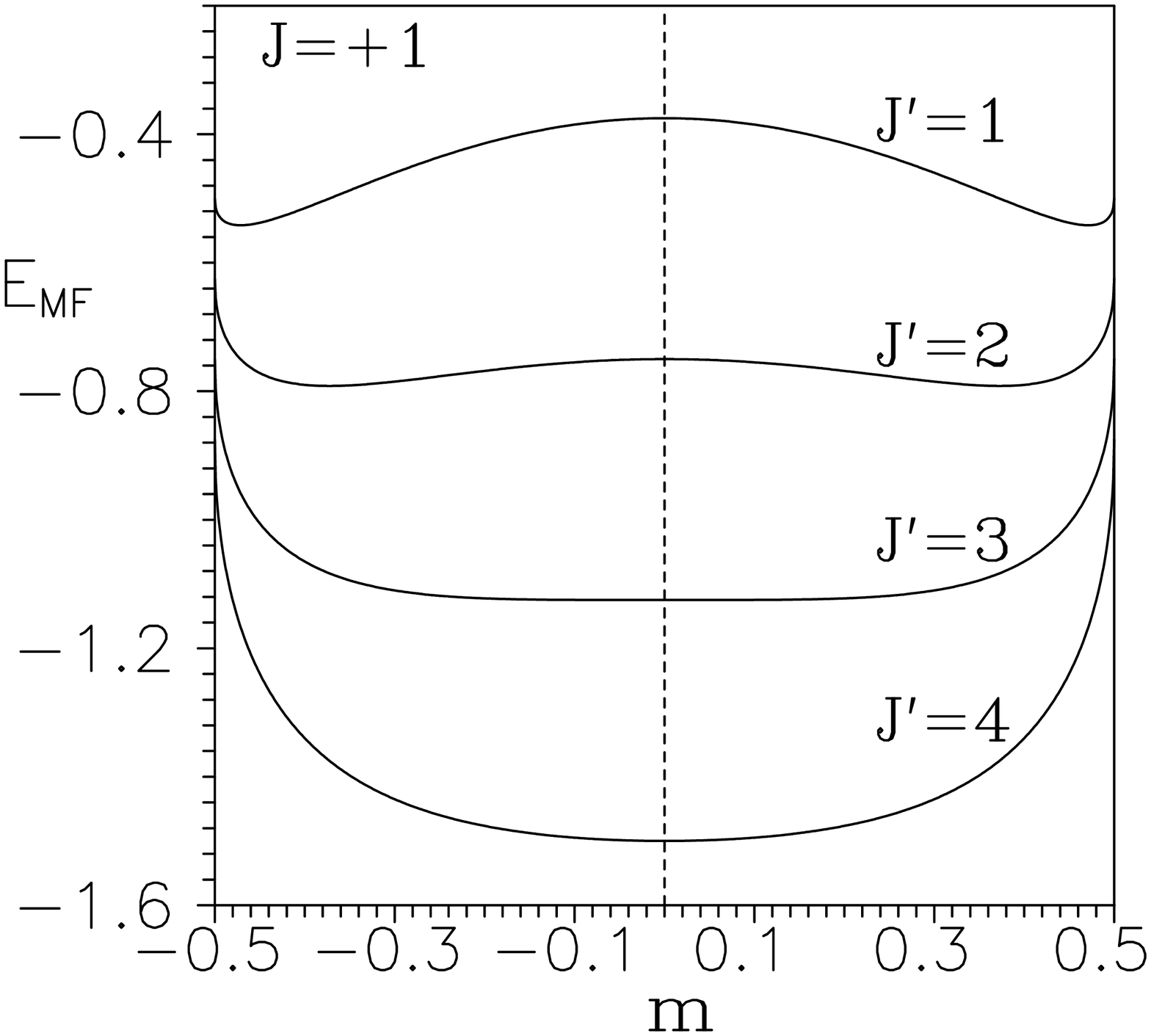}  % postscript image file name
\caption{Sublattice magnetization versus $J'$ (for $J=1$) (left graph)  
and energy versus sublattice magnetization (right graph) 
using a mean field approach 
(\ref{psi_mf}).
  \label{fig2}}
\end{figure}

{\it Mean Field Approach: }
We start with a simple mean-field (MF) like description of the
order-disorder transition.
The corresponding uncorrelated MF state for N\'eel LRO is the N\'eel state
$|\phi_{{\rm MF}_1}\rangle  = |\uparrow \downarrow \uparrow \dots \rangle$,
and for the dimerized singlet state it is
the rotationally-invariant 
product state of local pair singlets $|\phi_{{\rm MF}_2}\rangle  = 
\prod_{i \in A} [|\uparrow_{i} \downarrow_{i + \hat{x}}\rangle$
$-|\downarrow_{i} \uparrow_{i + \hat{x}}\rangle]/\sqrt{2}\;$ 
where 
$i$ and $i+\hat{x}$ correspond to those sites which cover the $J'$ bonds. 
In order to describe the transition between both states, we consider 
an uncorrelated
product state interpolating between $|\phi_{{\rm MF}_1}\rangle$ and 
$|\phi_{{\rm MF}_2}\rangle$ of the form\cite{gros_wenzel_richter,kruger00}
\begin{equation} \label{psi_mf} 
|\Psi_{\rm MF}(t)\rangle = \prod_{i \in A}
\frac{1}{\sqrt{1+t^2}} \left[|\uparrow_i \downarrow_{i+\hat{x}}\rangle
-t |\downarrow_i \uparrow_{i+\hat{x}}\rangle\right].
\end{equation}
We have 
 $ |\Psi_{\rm MF}(t=0)\rangle = |\phi_{\rm MF_1}\rangle$  and 
 $ |\Psi_{\rm MF}(t=1)\rangle = |\phi_{\rm MF_2}\rangle$.
We minimize $\langle\Psi_{\rm MF}|H| \Psi_{\rm MF}\rangle$ with respect to $t$
and obtain
\begin{equation}
\label{en_{MF}}
   \frac{E_{\rm MF}}{N} = 
\frac{ \langle\Psi_{\rm MF}|H| \Psi_{\rm MF}\rangle}{N} 
= \left\{\begin{array}{ll}
        -\frac{3J'}{8}-\frac{1}{24J}(3J-J')^2  & \quad J'\le 3 J\\
        -\frac{3J'}{8}            & \quad J'>   3 J,\\
    \end{array}\right.
\end{equation}
for the energy per site. For the sublattice magnetization 
$m\equiv\langle\Psi_{\rm MF}|S_{i \in A}^z|\Psi_{\rm MF}\rangle$ we get
$m=\sqrt{(3J-J')(3J+J')}/(6J)$ for $J'\le 3J$ and $m=0$ otherwise.
%
%Furthermore, it is found that the sublattice magnetisation has the
%following form,
%\begin{equation} \label{m_{MF}}
%  m=\langle\Psi_{\rm MF}|S_{i \in A}^z|\Psi_{\rm MF}\rangle = \left\{\begin{array}{ll}
%         \frac{1}{6J}\sqrt{(3J-J')(3J+J')} & \quad J'\le 3 J\\
%         0                  & \quad J'>   3 J.\\
%    \end{array}\right.
%\end{equation}
Note that $m$ vanishes at a critical point $J'_c=3J$, and that the critical 
index is the MF index $1/2$. 
Eq.~(\ref{en_{MF}}) may be rewritten in terms of $m$ as 
$ E_{\rm MF}/N  =      -\frac{1}{8}J'\hspace{0.1cm}-
\hspace{0.1cm}\frac{1}{4}J' 
\sqrt{1\hspace{0.1cm}-\hspace{0.1cm}4m^2} 
\hspace{0.1cm}-
\hspace{0.1cm} \frac{3}{2}J\hspace{0.1cm}m^2 \hspace{0.1cm}
$, and  Fig.~\ref{fig2} illustrates that the dependence of $E_{\rm MF}$ on $m$
corresponds to a typical scenario of  a second-order transition.  
We can expand $E_{\rm MF}$ up to the fourth order in $m$ near the critical point 
and find a Landau-type expression, given by 
$
 E_{\rm MF}/N  \;=\;  -\frac{3}{8}J'  
\hspace{0.05cm}+\hspace{0.05cm}\frac{1}{2} 
\left (J' - 3J\right) \hspace{0.05cm} m^2 
\hspace{0.05cm} +
\hspace{0.05cm} 
\frac{1}{2} J'\hspace{0.1cm}m^4 \;$.
However, as discussed elsewhere for a similar magnetic model for 
CaV$_4$O$_9$,\cite{troyer97}
MF theory probably does not describe the critical behaviour correctly.

\begin{figure}[t]
\epsfxsize=12.1pc % will enlarge or reduce the postscript figures based on the xsize
\epsfbox{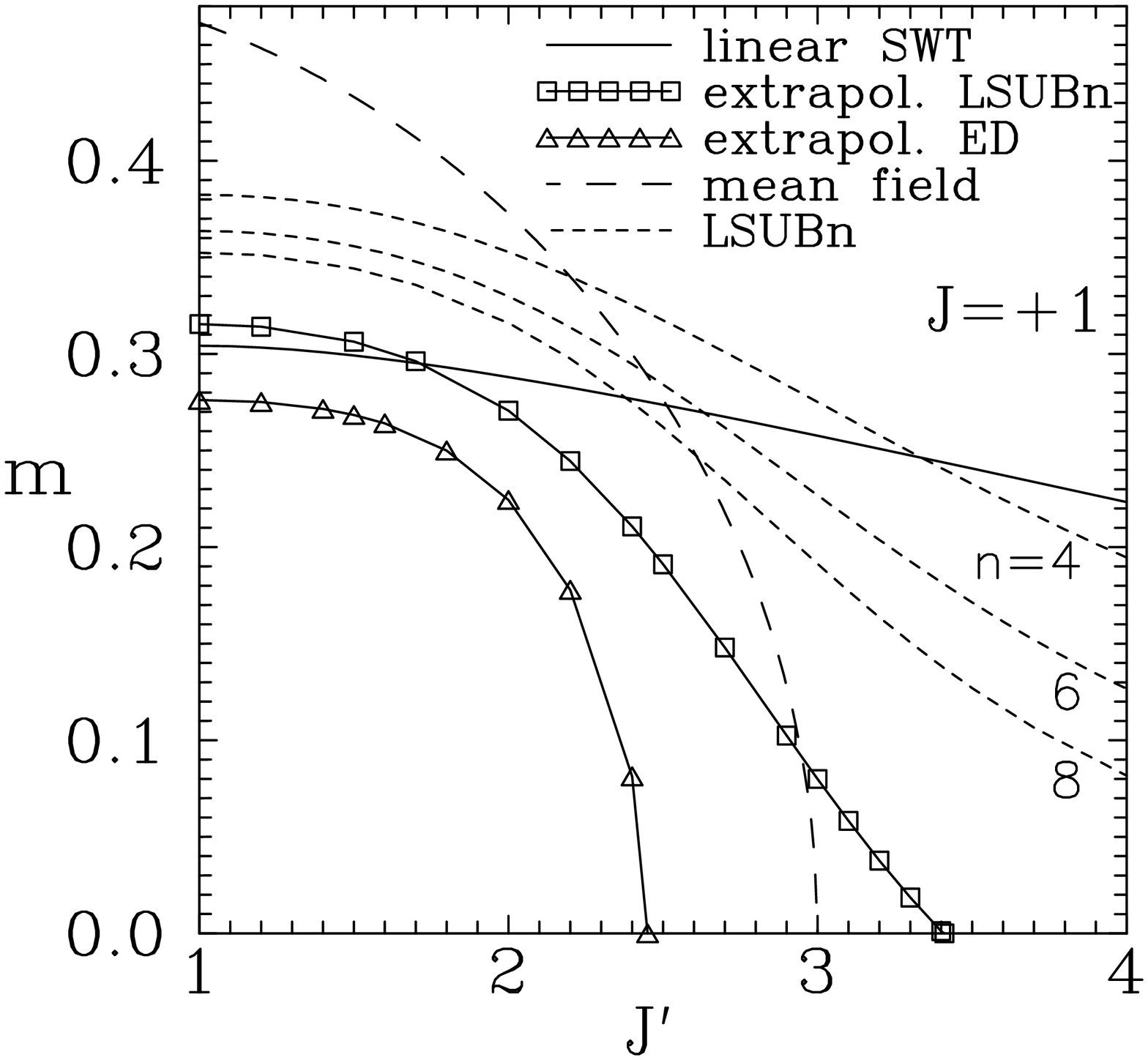}  % postscript image file name
\epsfxsize=12.1pc % will enlarge or reduce the postscript figures based on the xsize
\hspace{1cm} \epsfbox{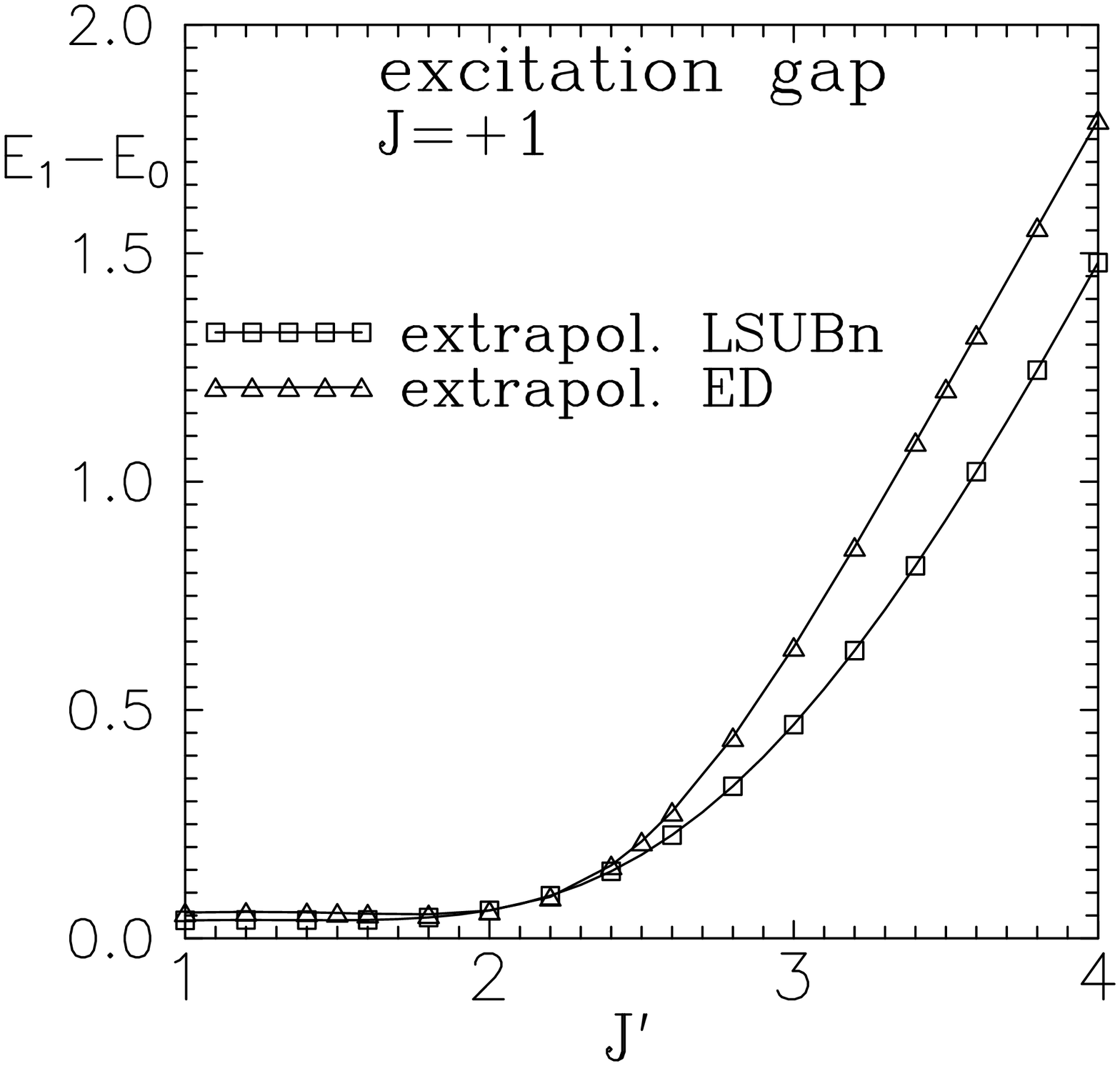} \caption{Sublattice magnetization (left graph) 
and excitation gap (right
graph) versus $J'$. 
  \label{fig3}}
\end{figure}

{\it CCM:}
Let us now apply a high-order CCM approach (for details see Refs.~\cite{zeng98,kruger00}) 
to this model. 
We set the classical collinear N\'{e}el state
to be the reference state $|\Phi\rangle$. We calculate the 
GS wave function, $|\Psi \rangle = e^S |\Phi\rangle$ within
the LSUB$n$ approximation scheme up to $n=8$ 
and extrapolate to $n \to \infty$. 
The CCM results for the order parameter are shown in the left graph of
Fig.~\ref{fig3} and they are compared to results  of linear spin wave theory (SWT), exact
diagonalization (ED) of N=16,18,20,26,32 sites, and the MF theory.
% KW SWT ED und data in Figure 
The CCM is able to describe correctly the order-disorder
transition, whereas conventional SWT  cannot
(for more details concerning the SWT and ED results see 
Ref.~\cite{kruger00}).
The critical value predicted by extrapolation of the LSUB$n$ results is, 
however, found to be slightly too large.
%%However, the extrapolation of LSUBn slightly overestimates the critical value.
We may also consider the inflection points of $m$ versus $J'$ for the 
LSUBn approximations. It is assumed that the true $m(J')$-curve will have 
a negative curvature up to the critical point. Thus 
we might expect that (for increasing $n$) 
the
inflection point approaches the critical point. We find the corresponding 
inflection
points  at $J'=3.1$ (n=2), $J'=3.0$ (n=4), 
 $J'=2.9$ (n=6) and  $J'=2.85$ (n=8), indicating a critical
value $J'_c$ somewhere between $2.5J$ and $3J$. 
Notice, that the estimation of $2.5 \lesssim J_c'/J \lesssim 3$ 
is consistent with results of series expansions and exact
diagonalizations.\cite{singh88,kruger00}
The breakdown of N\'eel LRO due to singlet formation is also accompanied by the
opening of an excitation gap between the singlet GS and the
first triplet excitation. This behaviour is well described by the CCM 
(right graph of Fig.~\ref{fig3}) which predicts that the 
gap opens in the range $2J < J' < 3J$
(and notice that the non-zero gap below $2J$ is a result of the limited 
accuracy of the extrapolation).

\section{Competition with Frustration}
We now consider the frustrated model (where $J$ and $J'$ have different
signs). Due to the incommensurate classical spiral state the
ED technique for finite-size systems is less appropriate. 
The CCM intrinsically considers the limit $N \to \infty$ from the outset 
and thus has no problems in dealing
with incommensurate states. Hence the CCM 
appears to be particularly suitable to 
attack the frustrated quantum model.
We choose the classical state to be our CCM model state,
although quantum fluctuations may change the pitch angle
of the spiral phase.
Hence, we determine the `quantum pitch angle' $\Phi$ by minimizing 
$E_{{\rm LSUB}n}(\Phi)$ with respect to $\Phi$.
\begin{figure}[t]
\epsfxsize=12.1pc % will enlarge or reduce the postscript figures based on the xsize
\epsfbox{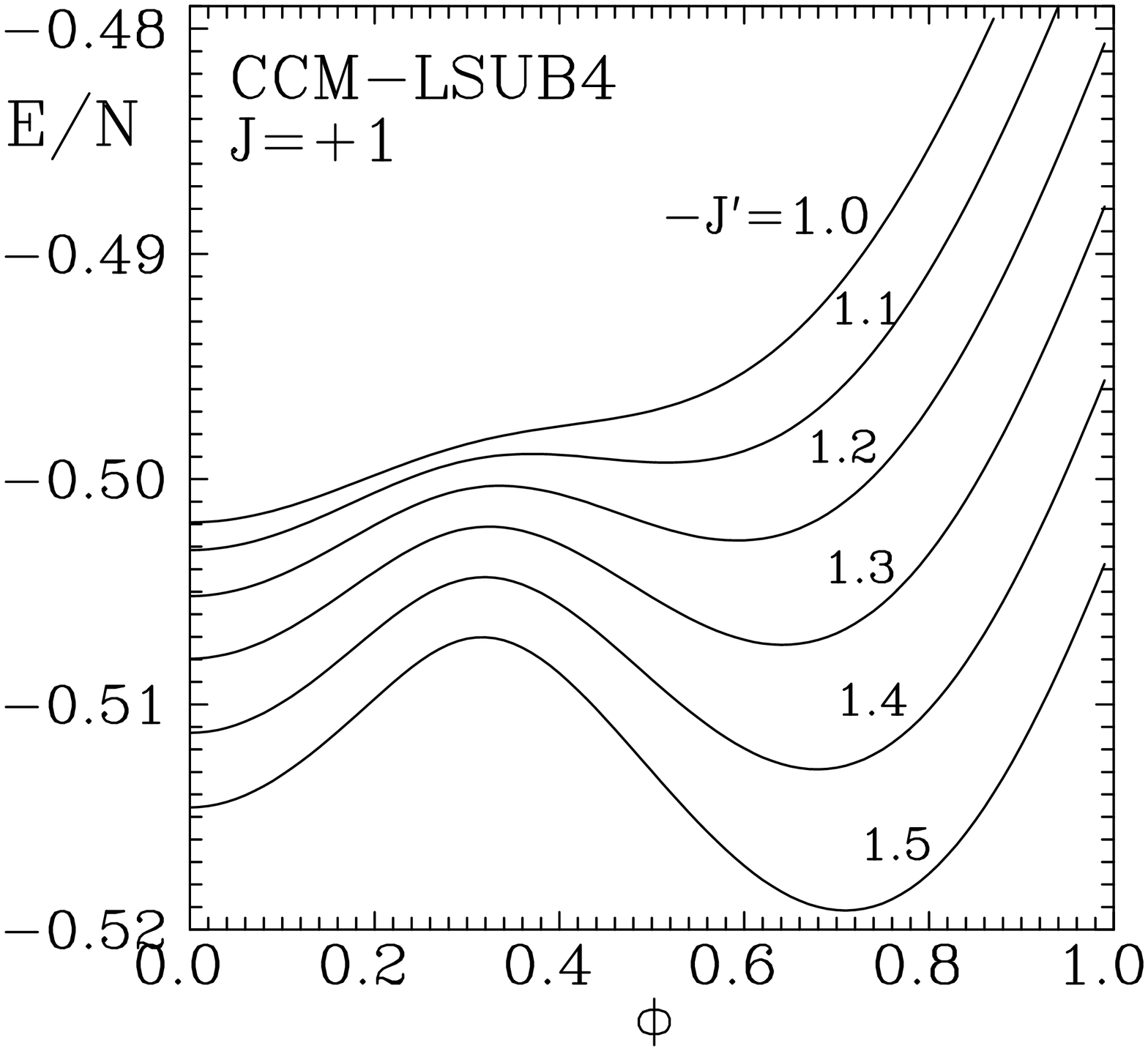} \epsfxsize=11.9pc \hspace{1cm}
\epsfbox{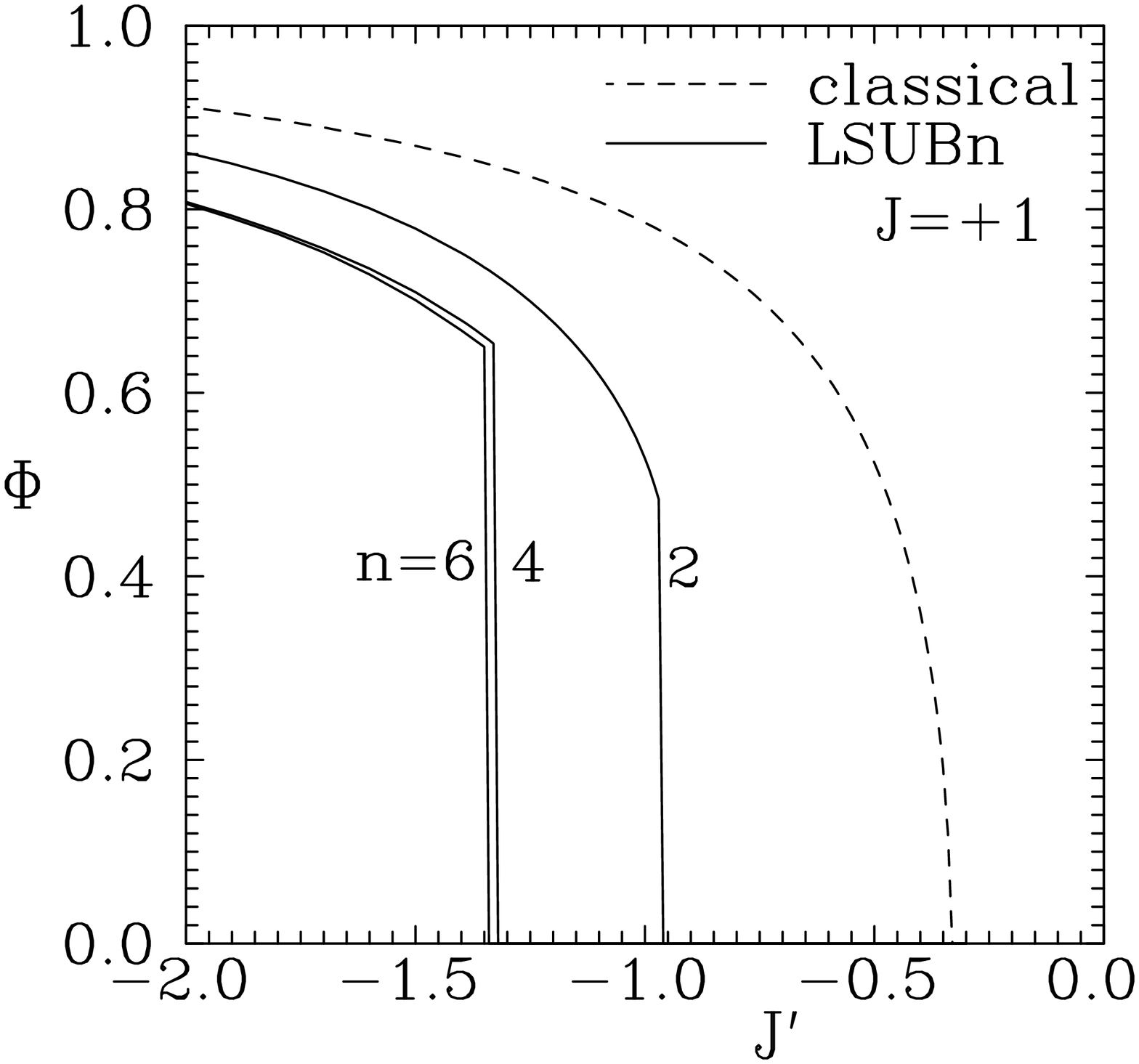} % postscript image file name
\caption{Energy versus quantum pitch angle for LSUB4 (left graph) and 
quantum pitch angle versus $J'$ (right graph). Note, that $\Phi=0$
corresponds to the N\'eel state.  
  \label{fig4}}
\end{figure}
\begin{figure}[t]
\epsfxsize=12.5pc % will enlarge or reduce the postscript figures based on the xsize
\epsfbox{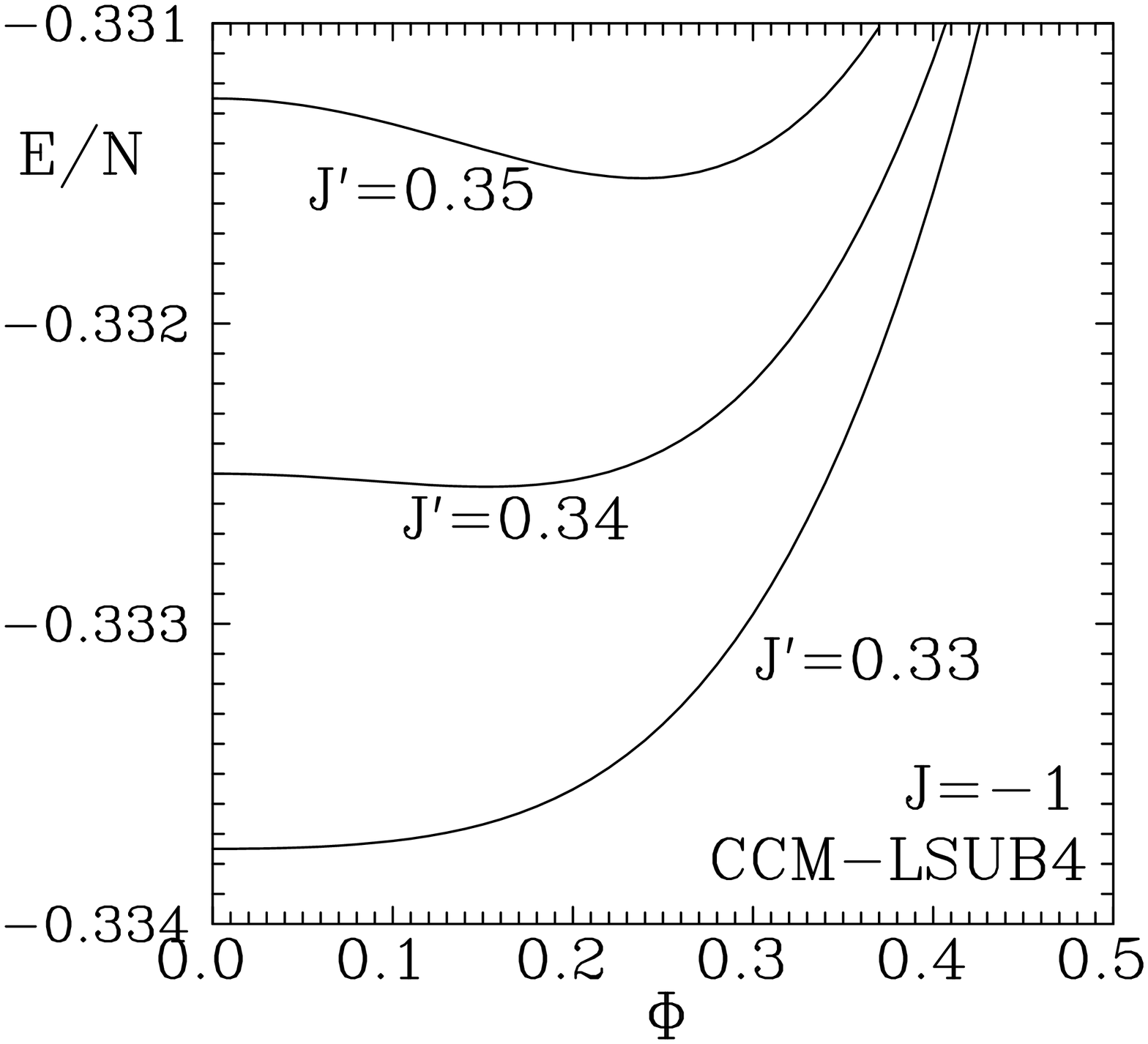} \epsfxsize=11.9pc \hspace{1cm}
\epsfbox{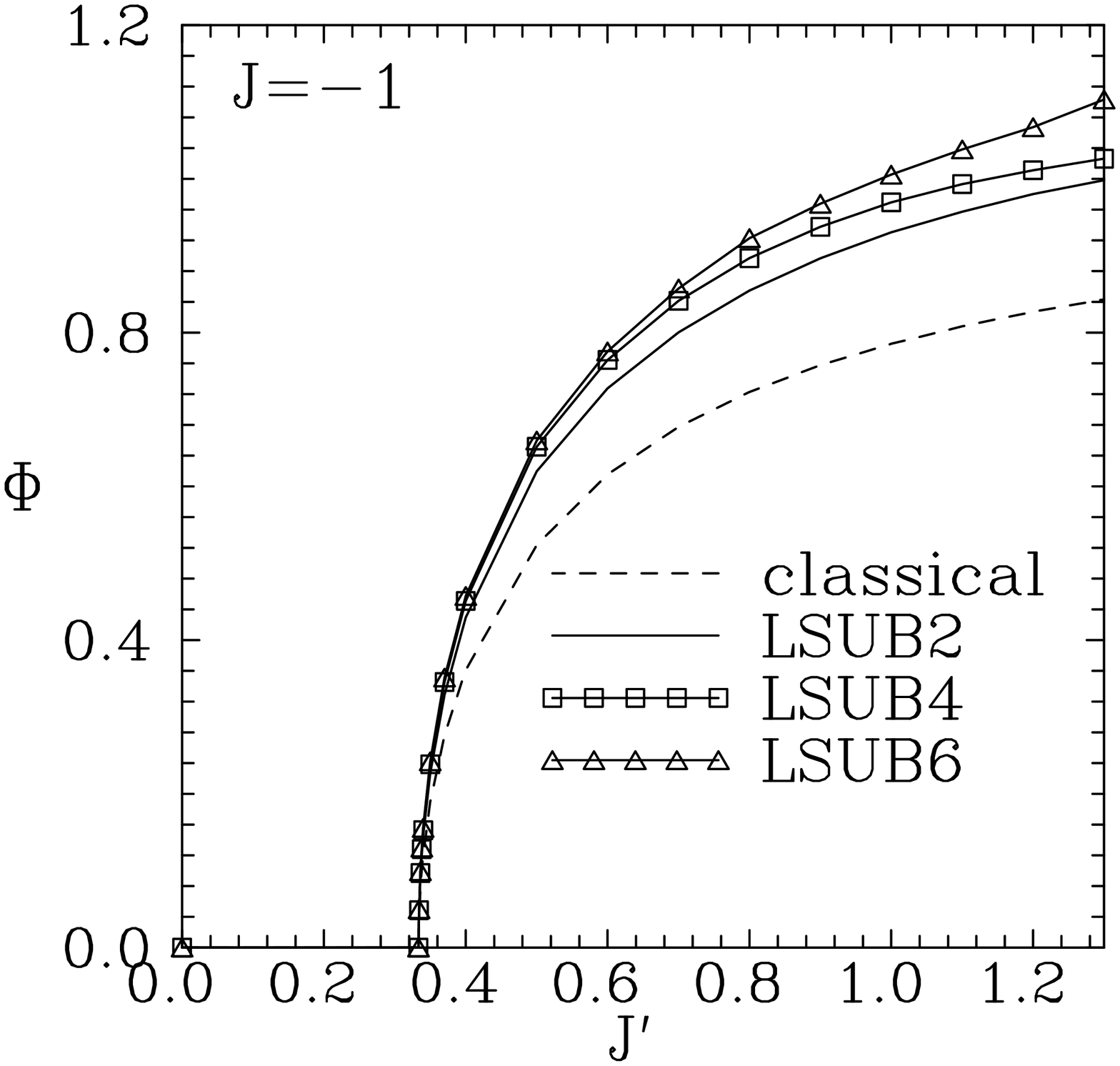} % postscript image file name
\caption{Energy versus quantum pitch angle for LSUB4 (left graph) and 
quantum pitch angle versus $J'$ (right graph). Note, that $\Phi=0$
corresponds to the fully polarized ferromagnetic state.  
  \label{fig5}}
\end{figure}

{\it N\'eel versus Spiral:}
We consider (antiferromagnetic) $J=+1$ and (ferromagnetic) $J' <
0$. Results for $E( \Phi) $  and $\Phi(J') $ are shown in Fig.~\ref{fig4}.
The main results are that: (i) In the quantum case the quantum 
N\'eel state 
remains the GS up to much stronger frustration than in the
classical case. Indeed, it is generally found for spin
systems, that quantum fluctuations favour collinear spin structures as
opposed to 
noncollinear ones.   
(ii) The quantum fluctuations change the phase transition from second order
to first order.   
(iii) The CCM yields a consistent description of the collinear and the spiral 
phases.
   
{\it Ferro versus Spiral:}
%As discussed above quantum fluctuations can strongly 
%affect zero temperature transitions. 
%To support this statement 
We now consider the model for 
(ferromagnetic) $J=-1$ and (antiferromagnetic) $J' >0$. 
In the classical model we again have a second-order 
transition from a collinear to a spiral state (see section \ref{cgs}).
In the quantum model the situation is quite different. Although the
collinear antiferromagnetic state possesses strong quantum fluctuations,
the collinear, fully polarized, ferromagnetic state possesses no such 
quantum fluctuations.
The corresponding 
results for $E( \Phi) $  and $\Phi(J') $ are shown in Fig.~\ref{fig5}.
By contrast to the situation at 
$J=+1$, the transition from the ferromagnetic to the noncollinear
spiral is now of second order, the same as for the classical model. 
Furthermore, the classical critical point $J'/J = -1/3$ also holds for the 
quantum case.
\begin{figure}[t]
\epsfxsize=10.9pc % will enlarge or reduce the postscript figures based on the xsize
\epsfbox{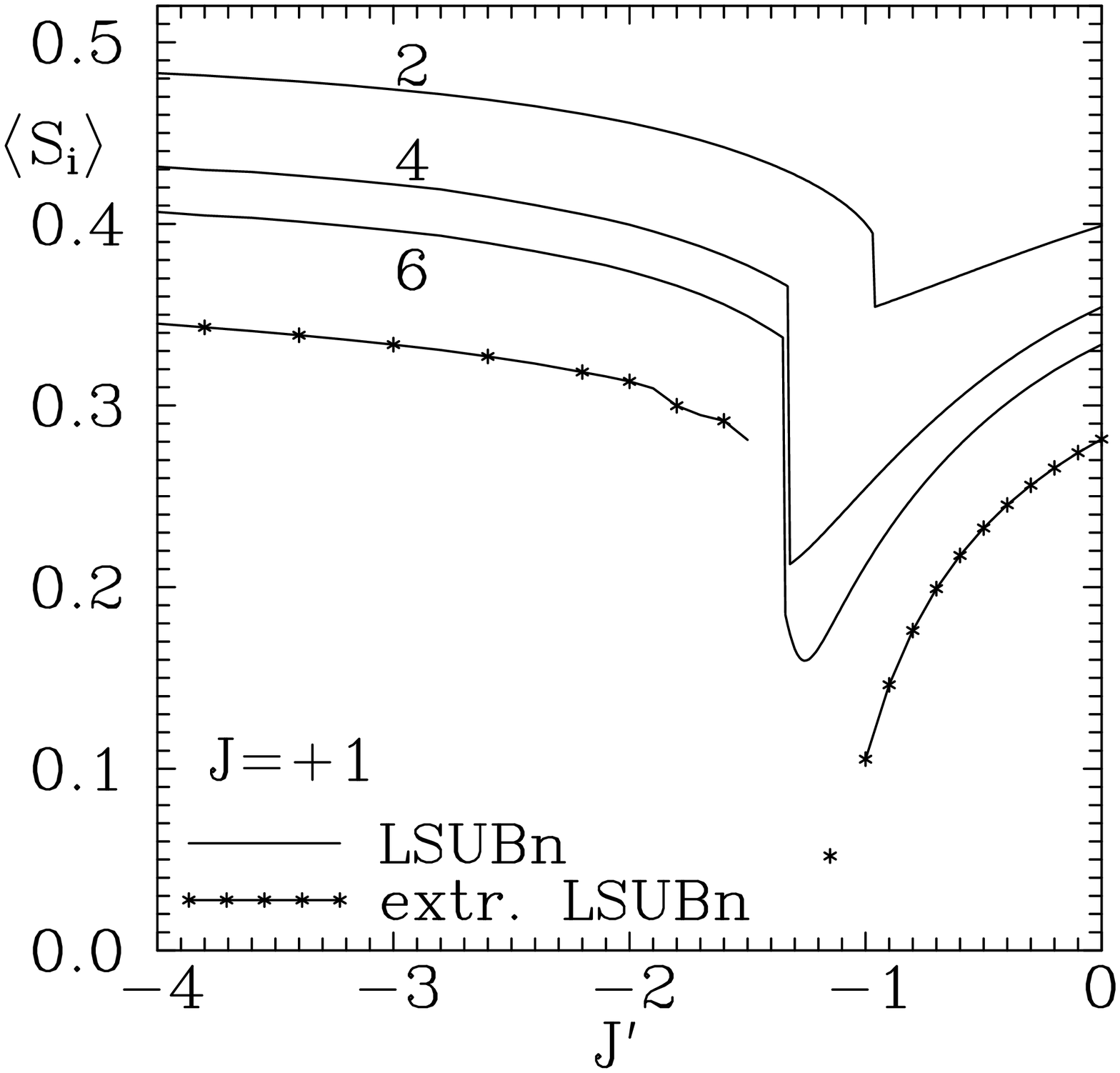} \epsfxsize=10.9pc \hspace{1cm}
\epsfbox{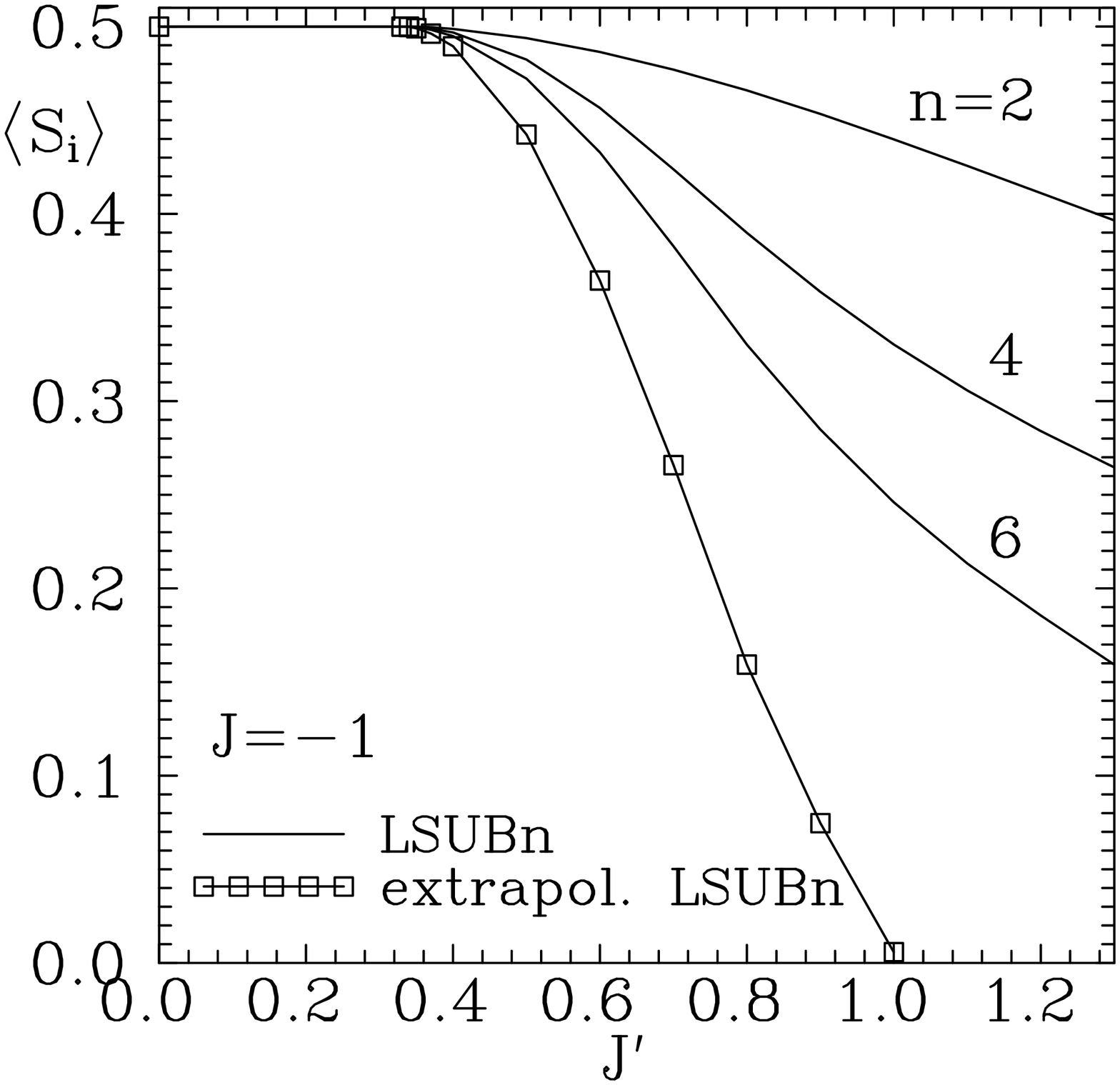} % postscript image file name
\caption{On-site magnetic moment versus $J'$ for $J=+1$ (left graph) and
$J=-1$ (right graph).
  \label{fig6}}
\end{figure}
The difference between both  cases also becomes evident when the
order parameter $\langle S_i \rangle$ is considered (Fig.~\ref{fig6}). 
For $J=+1$ there is
a discontinuity in $\langle S_i \rangle$ at every level of
LSUBn approximation. However, the extrapolation to $n \to
\infty$ becomes imprecise close to the phase transition point. 
We cannot therefore decide whether
the order parameter vanishes near to the transition point.
For $J=-1$ there is
a smooth change  in $\langle S_i \rangle$ at the critical point.
Increasing the antiferromagnetic $J'$ the spiral magnetic
order becomes weaker and vanishes at $J' \approx 1$. The underlying
reason for that is local singlet formation, as discussed in section
\ref{nofr}, and the continuous vanishing of the spiral order is 
therefore very similar to
this second-order transition. However, the strength of $J'$ 
needed for local singlet formation is
much smaller due to the assisting effects of frustration 
(c.f., Ref.\cite{gros_wenzel_richter}).

\section{Conclusions}
In this article we have investigated the zero-temperature phase 
transitions 
of a spin-half Heisenberg
system on the square lattice. The main results of our treatment are:
(i) Quantum fluctuations plus competition without frustration are able to
destroy N\'eel LRO by local singlet formation. This is a pure quantum effect
and has no classical counterpart. The control parameter is the strength of
the competition and the breakdown of N\'eel ordering is
accompanied by the opening of a spin gap. Standard SWT
(even to higher orders)
fails to describe this transition, whereas the CCM describes both 
the order parameter and
the gap satisfactorily. Since we have no frustration, most standard
techniques (e.g., QMC) are applicable and a quantitative
description is possible. As was discussed in Ref.\cite{troyer97}, the
critical properties seem to correspond to the 3D classical Heisenberg
model.
(ii) Competition due to frustration was found to give more complex 
magnetic properties. In the model considered 
we have a second-order transition between collinear
(antiferro- or ferro-magnetic) and noncollinear (spiral) states
 driven by frustration in the classical case. 
In the quantum spin-half model, standard
techniques (e.g., QMC) are not applicable due to the violation of Marshall's sign
rule. By contrast, the CCM provides a consistent description of collinear,
noncollinear, and disordered phases. Furthermore, we find a strong influence of quantum
fluctuations on the nature of the collinear-noncollinear transition, and
quantum fluctuations (which favour collinear ordering) may change the
second-order classical transition to a first-order quantum transition. If
quantum fluctuations are suppressed in the collinear phase, the 
transition to the spiral phase is similar for the quantum and
classical models.  

\section*{Acknowledgments}
We thank the Deutsche Forschungsgemeinschaft 
(Ri 615/9-1) for its support.


\begin{thebibliography}{99}
\bibitem{qpt} S.~Sachdev, {\em Quantum Phase Transitions}
                   (Cambridge University Press 1999).
\bibitem{mermin66}
     N. Mermin and H. Wagner, \Journal{Phys. Rev. Lett.}{17}{1133}{1966}.
\bibitem{manou91}
    E. Manousakis, \Journal{Rev. Mod. Phys.}{63}{1}{1991}.
\bibitem{anderson73}
    P.W. Anderson, \Journal{Mater. Res. Bull.}{8}{153}{1973};
    P. Fazekas and P.W. Anderson,  \Journal{Phil. Mag.}{30}{423}{1974}.
\bibitem{MP_rule}
 J. Richter, N.B. Ivanov, and R. Retzlaff,
      \Journal{Europhys. Lett.}{25}{545}{1994}.
%A.~Voigt,  J.~Richter,  and N.B.~Ivanov, 
%      \Journal{Physica A}{245}{269}{1997}.
\bibitem{ri93}
   J. Richter, \Journal{Phys. Rev. B}{47}{5794}{1993}.
\bibitem{oitmaa96} J. Oitmaa and Zheng Weihong, 
              \Journal{Phys. Rev. B}{54}{3022}{1996}.
\bibitem{Bishop6} R.F. Bishop, D.J.J. Farnell, and J.B. Parkinson,
  \Journal{Phys. Rev. B}{58}{6394}{1998}.  % J1-J2-model mit CCM

\bibitem{sorella00} L.~Capriotti and S.~Sorella, \Journal{Phys.Rev.Lett.}{84}{3173}{2000}. 

\bibitem{xian95} Y. Xian, \Journal{Phys. Rev. B}{52}{12485}{1995}.
\bibitem{niggemann97} H.~Niggemann, G.~Uimin, and J.~Zittartz,
              \Journal{J.Phys.: Condens. Matter}{9}{9031}{1997}.
\bibitem{richter98} J~.Richter, N.B.~Ivanov, and J.~Schulenburg,
              \Journal{J.Phys.: Condens.Matter}{10}{3635}{1998}.
\bibitem{koga00} A.~Koga et al., \Journal{Phys. Rev. B}{62}{5558}{2000}.

\bibitem{sandvik_scalapino} A.W. Sandvik and D.J. Scalapino,
                            \Journal{Phys. Rev. Lett.}{72}{2777}{1994}.
\bibitem{gros_wenzel_richter}
    C. Gros, W. Wenzel, and J. Richter,
    \Journal{Europhys. Lett.}{32}{747}{1995}.


\bibitem{troyer96}  M. Troyer, H. Kontani, and K. Ueda,
                       \Journal{Phys. Rev. Lett.}{76}{3822}{1996}.
\bibitem{troyer97}  
                    M. Troyer, M. Imada, and K. Ueda,
                       \Journal{J. Phys. Soc. Jpn.}{66}{2957}{1997}.

%\bibitem{kageyama99}
%    H.Kageyama, K.Yoshimura, R.Stern, N.V.Mushnikov, K.Onizuka, M.Kato,
%    K.Kosuge, C.P.Slichter, T.Goto, Y.Ueda, 
%   \Journal{Phys.Rev.Lett.}{82}{3168}{1999}.

\bibitem{kageyama99} H.~Kageyama et al., \Journal{Phys. Rev.
Lett.}{82}{3168}{1999}.

\bibitem{koga00a} A.~Koga and N.~Kawakami, \Journal{Phys. Rev.
Lett.}{84}{4467}{2000}.

%%% general CCM
%\bibitem{coester58} F. Coester,
%                \Journal{Nucl. Phys.}{7}{421}{1958};
%                F. Coester and H. K\"ummel, 
%                 \Journal{Nucl. Phys.}{17}{477}{1960}.
\bibitem{bishop91} R.F.~Bishop in {\em Microscopic Many-Body Theories and
                   Their Applications}, eds. J.~Navarro and A.~Polls,
                   Lecture Notes in Physics Vol.~{\bf 510} (Springer
                   1998).
%% previous CCM spin lattice
%\bibitem{roger90} M. Roger and J.H. Hetherington,
%                \Journal{Phys. Rev. B} {41}{200 }{1990}.
\bibitem{bishop91a} R.F. Bishop, J.B. Parkinson, and Y. Xian,
                 \Journal{ Phys. Rev. B} { 43}{ 13782} {1991};
                 \Journal{Phys. Rev. B} { 44}{ 9425} {1991}.
\bibitem{Bishop2} R.F. Bishop, R.G. Hale, and Y. Xian,
                 \Journal{Phys. Rev. Lett.}{73}{3157}{1994}.
\bibitem{zeng98} C. Zeng, D.J.J. Farnell, and R.F. Bishop,
                 \Journal{J. Stat. Phys.}{90}{327}{1998}.
\bibitem{bishop99} R.F. Bishop, D.J.J. Farnell, and C. Zeng,
                 \Journal{Phys. Rev. B}{59}{ 1000} {1999}.  % nodal surfaces
\bibitem{kruger00} S.E.~Kr\"uger, J.~Richter, J.~Schulenburg,
  D.J.J. Farnell, and R.F. Bishop, 
                  \Journal{Phys. Rev. B} { 61}{ 14607} {2000}.  % J-J' CCM
\bibitem{singh88}
    R.R.P.~Singh, M.P.~Gelfand, and D.A.~Huse,
    \Journal{Phys. Rev. Lett.}{ 61}{ 2484 }{1988}.

\end{thebibliography}
\end{document}